# Analyzing Partitioned FAIR Health Data Responsibly

Chang Sun[a], Lianne Ippel[a], Birgit Wouters[a], Johan van Soest[b], Alexander Malic[a], Onaopepo Adekunle[a], Bob van den Berg[c], Marco Puts[c], Ole Mussmann[c], Annemarie Koster[d], Carla van der Kallen[e], David Townend[f], Andre Dekker[b], Michel Dumontier[a]

[a] *Institute of Data Science, Maastricht University, The Netherlands*
[b] *Department of Radiation Oncology (MAASTRO), GROW School for Oncology and Developmental Biology, Maastricht University Medical Centre+, The Netherlands*
[c] *Centraal Bureau voor de Statistiek (CBS), The Netherlands*
[d] *Department of Social Medicine, CAPHRI Care and Public Health Research Institute, Maastricht University*
[e] *Department of Internal Medicine, CARIM School for Cardiovascular Diseases, Maastricht University*
[f] *Department of Health, Ethics and Society, CAPHRI Research School, Maastricht University*

*It is widely anticipated that the use of health-related big data will enable further understanding and improvements in human health and wellbeing. Our current project, funded through the Dutch National Research Agenda, aims to explore the relationship between the development of diabetes and socio-economic factors such as lifestyle and health care utilization. The analysis involves combining data from the Maastricht Study (DMS), a prospective clinical study, and data collected by Statistics Netherlands (CBS) as part of its routine operations. However, a wide array of social, legal, technical, and scientific issues hinder the analysis. In this paper, we describe these challenges and our progress towards addressing them.*

## 1. INTRODUCTION

Using "Big Data" in a responsible manner is essential to secure the trust needed to make sensitive data available for addressing some of the most challenging problems of our time. In particular, the use of big data for health offers unprecedented opportunities to improve our understanding of diseases, to augment diagnostic accuracy and efficiency [1, 2], and to facilitate the transition to precision medicine [3, 4]. However, a major barrier to realizing these and other applications is that health-related data are dispersed widely in both their location (e.g. medical records, consumer activity, social media), representation (structured, semi-structured, and unstructured), and in who is responsible for their collection and governance (individuals, health care providers, insurers, government agencies). This distributed nature of the health data, unlike that seen in many other domains, poses substantial challenges to use health data in understanding and improving the health and wellbeing of individuals.

The health and wellbeing of individuals and communities arise from a complex combination of social and economic factors including where we live, the quality of our environment, the income we earn, our level of education, the strength of our relationships with friends and family, as well as access and use of health care services [5]. Understanding the nature and significance of social and economic determinants of health can provide governments and policy-makers a basis to pursue effective policies to improve health outcomes while minimizing health expenditures. However, gaining an understanding of the nature of these relationships can be explored through statistical analysis over big data with plenty of limitations. A wide variety of social, legal-ethical, political, technical, and scientific questions must be addressed to ensure the responsible use of these data for health research. For instance, have patients provided consent for their data to be used for research purposes? Have they consented to have their data combined with other, non-research data? Is it necessary to obtain a new consent for this further processing of already collected data, or is it compatible with the legal basis for the original processing of these data? Are public institutions such as Statistics Netherlands (CBS) obligated to safeguard data they collect about citizens in a manner that precludes their automatic discovery and reuse? What scientific methods are most appropriate for analyzing personal data collected from two completely different data sources? Can we learn health determinants without compromising individual privacy? These, and many other questions must be addressed for health-related data to be used in a responsible (and effective) manner.

## 1.1. OBJECTIVES

The overall goal of our project is to establish a scalable and responsible framework to uncover the determinants of human health using access-restricted data from multiple parties with different data governance policies in a privacy-preserving manner. In our project, 'the privacy-preserving manner' guarantees 1) no personal identifiable data is revealed to any party; 2) no data party can see other parties' data; and, 3) results of data analysis do not leak any sensitive or identifiable information. We focus our research to studying the relationship between diabetes and socio-economic factors such as lifestyle and health-care utilization using data from the Maastricht Study (DMS), a prospective clinical study, and from Statistics Netherlands (CBS), a government agency. Our approach involves the development of three work packages: scientific (research question and methods); technical (architecture and software); and social, legal, and ethical aspects. Our work, funded as part of the Dutch National Research Agenda, aims to create new social and technical infrastructure to enable the responsible use of big data in research.

## 1.2. CHALLENGES

From a technical perspective, as the data from the same cohort are distributed at CBS and the Maastricht Study, a first challenge is how to link the individual data between two data entities that are sensitive, and access-restricted. While linking two datasets through a common identifier, such as the Dutch national identifier (BSN), is a natural first thought, legal restrictions prohibit this approach, even for research purposes. This topic is not properly addressed yet in the literature [6]. A second challenge is whether an association can be learned from distributed variables without revealing sensitive data to any party. Lastly, there is a question of how to evaluate: 1) the security; 2) to what extent privacy is preserved; and 3) the technical performance (e.g., accuracy, scalability, and stability) of the infrastructure.

From a social, legal, ethical perspective, a key legal issue is the General Data Protection Regulation 2016/679 (GDPR), which is the EU legislation providing new rights and responsibilities regarding personal data that came into effect on 25th May, 2018, with direct effect in Member States' law. While differences between the former Data Protection Directive 95/46/EC (DPD) and the GDPR appear small, they are sufficiently significant to require a new interpretation. Further, the shift to a Regulation with direct effect would imply that the problems of harmonisation under the old law would be very much overcome. However, the GDPR contains many places where there are explicit choices for Member States, and other places where the concepts and language of the law are unclear. Therefore, it remains important to observe not only the difference between the old and new law, but also how Member States' Supervisory Authorities and how Member States' laws on data protection interpret the GDPR, and how the EU itself responds to these differences. . Depending on these interpretations, the GDPR could potentially facilitate or hinder Big Data research. A balance must be found in that interpretation to ensure both individual privacy (particularly through effective respect for confidentiality) and solidarity (to underpin health research). To achieve this, emerging data science tools for effective security and governance must also be explored alongside book-based law and ethics solutions; further, the balance must resonate with the sensitivities expressed in the general public around the governance and use of personal data in Big Data health research.

## 2. METHODS

The project is divided into three interlocking work packages (WP). The Scientific WP focuses primarily on methods for statistical learning from big data, to explore the interesting association between social and economic factors and type 2 diabetes. The Technical WP aims to develop a computational framework to facilitate access, reuse, and combining data from the Maastricht Study and CBS in a secure environment, essentially by addressing the FAIR principles[1]. The Ethics, Law, and Societal Issues (ELSI) WP explores governance frameworks for linking sensitive data, including the legal and ethical basis for the procuring and processing of data, and then the development a broader and scalable governance that will define and underpin the responsible use of Big Data in the health domain.

### 2.1 SCIENTIFIC WORK PACKAGE

The scientific, technical, and ELSI WPs explore the means to share relevant data between two data providers, the Maastricht Study and CBS. In order to explore the potential associations between two distributed datasets, it is essential that the datasets have 1) a substantial set of the individuals in common, and 2) at least some variables in common to allow the datasets to be linked. The scientific WP, in combination with the ELSI group, investigates relevant and appropriate variables at both data sites, which are of high quality (i.e., accurate and complete) and are allowed to be used for scientific research according to participants' legal consents. At this stage, we requested data regarding the *hospitalization*, *declared healthcare costs through basic health insurance,* and *medication use* from CBS. From the Maastricht Study, we requested data regarding *lifestyle, physical activity, diabetes status, physical function (SF36),* and *education*. Using these data, the scientific WP will explore the associations between physical activities and health status of different stages of Type 2 diabetes patients. The aim of the scientific WP is twofold: 1) tackle the health-related research question; 2) test and evaluate the performance of the new privacy-preserving infrastructure developed by the technical WP, which is discussed in Section 2.3. Once this infrastructure is proved to be sufficiently privacy-preserving and reliable from the legal-ethical and the technical perspective, many more scientific questions could potentially be addressed by the scientific WP.

---

[1] FAIR Principle: Findable, Accessible, Interoperable, and Reusable [7].

## 2.2. Ethical and legal Work Package

Ensuring public trust is an essential prerequisite in the conduct of big data research, particularly in light of recent compromises of trust and security of online information systems [8]. When public trust is secured, the great potential of these (sensitive) data to address the challenging health research questions can be explored. The primary goal of the ELSI WP is to ensure the responsible use of data in a privacy-preserving manner, whilst facilitating big data research. The facilitation of big data research is facing challenges, for instance by the implementation of the GDPR, giving its citizens more control and rights over their data, and by national laws complicating the matter of processing identifying information. Therefore there are three elements to the ELSI work: 1) to ensure that the work complies with law and ethics as they are commonly understood; 2) to explore how the novel work challenges the common understanding of the law and ethics, and to develop robust interpretations where law and ethics are uncertain; 3) to ensure that the proposed interpretations resonate with the sensitivities expressed by the general public, and seek to present the interpretations to regulatory authorities. At its heart, this ELSI work concerns developing an understanding of the "privacy-preserving" concept.

Whereas the GDPR seeks to harmonise EU law on the processing of personal data, it leaves a number of issues unclear and a number of issues for Member State discretion and choice. These uncertainties present issues for the ELSI work package. A good example of such a legal complication is the use of national identification numbers. Article 87 of the GDPR leaves it up to the national governments to determine the use of their national identification number. The Netherlands has adopted a very restrictive approach regarding the use of the national identification number (Burgerservicenummer - BSN): only certain situations allow for the use of the BSN, for example uses by governmental entities, by health insurers, and in matters related to tax. However, this reliable identifier cannot be used for scientific purposes. Of course, the trustworthiness and validity of analytic results produced by a Big Data project such as ours would benefit from the unambiguous mapping of personal data from the Maastricht Study with personal data collected by CBS by the use of the BSN, but it is not allowed. To achieve compliance with Dutch law, we will match the data using other variables that are common to both datasets.

Independent of which identifying information is used to link the data sets, this information is sensitive and should be treated carefully. To avoid breaching the privacy of the individuals, we need to pseudonymize all personal data[2]. Pseudonymization entails replacing the identifying fields with one or more artificial identifiers, or pseudonyms. To effectively pseudonymize data, Article 4 of the GDPR stipulates that the personal data must be processed such that personal data can no longer be attributed to a specific data subject, without the use of additional information which must be kept separately. Our compliance with this requirement is discussed in more detail below in Section 3.

Since the highly reliable identifiers - BSN is unavailable to applied in our project, an additional challenge lies in linking of differently sized datasets. CBS holds data on all Dutch citizens and residents, while the Maastricht Study only consists of a few thousand individuals. This difference of the size of the datasets hinders the linking of the datasets: the combination of variables used to match the records between the datasets might not result in unique matches. Depending on what non-national identifier fields are used for matching, a single record of the

---

[2] GDPR Article 4(1): " 'personal data' means any information relating to an identified or identifiable natural person ('data subject'); an identifiable natural person is one who can be identified, directly or indirectly, in particular by reference to an identifier such as a name, an identification number, location data, an online identifier or to one or more factors specific to the physical, physiological, genetic, mental, economic, cultural or social identity of that natural person."

Maastricht Study might match with multiple records from the CBS. It was, therefore, essential to limit the 'pool of potential matches' from CBS. Using the sample characteristics of the Maastricht Study, we can substantially limit the selection of data subjects of CBS: the Maastricht study is only open to participants between the age of 40 and 75, who live in the South of Limburg. However, limiting the CBS's selection increases the probability of (in)directly re-identification of a data subject. Only the collaboration with the other two WPs will ensure the strength and the effectiveness of the infrastructure and that the data science tools used will guarantee GDPR compliance. This collaboration exclusively will assure privacy-preserving research so as to strengthen the public trust.

As the project progresses, other legal challenges will emerge. While some argue that the GDPR provisions are incompatible with big data research [9], others argue that the GDPR might allow citizens to benefit from enhanced data protection, while enjoying the innovations enhanced data analytics bring [10]. It is this contradicting interpretation of the GDPR, situated at both ends of the spectrum, that makes the creation of the legal framework for this project so very complex. The ELSI Work Package has already spent time working on the interpretation of "compatible processing" (allowing secondary processing of already gathered data - our project's situation - because it is compatible with the original legal basis for the first processing of the data). This is an area where the GDPR is unclear in its drafting, and conceptual certainty must be found through both doctrinal (book-based) analysis, and public engagement work. In time, the European Data Protection Supervisor and the National Supervisory Authorities will find robust interpretations of the unclear parts of the GDPR. One of the aims of the research is to contribute to the construction of interpretations that clarify the uncertainties; our work seeking appropriate interpretations of unclear law depends on seeking approval of our ELSI research findings with the appropriate regulatory institutions.

## 2.3. TECHNICAL WORK PACKAGE

Given the sensitive nature of the data used in this project, the infrastructure should maximally safeguard the privacy of the data subjects. The technical WP is developing an infrastructure based on the Personal Health Train (PHT) architecture [11]. The PHT is designed to facilitate authorized algorithmic processing in a secure and privacy-preserving manner at multiple sites without requiring transfer of data to a centralized location. This eliminates the need to make multiple copies of the data, which are then difficult to track and audit any access to them. The PHT architecture comprises algorithm 'trains' that visit 'stations', which check algorithmic processing credentials and provide access to data it is authorized to release. Three PHT 'stations' will be established for this purpose: one at CBS, one at the Maastricht Study/Maastricht University, and one Trusted Secure Environment (TSE) station. In our project, the TSE is a shielded and isolated environment within CBS. Moreover, the 'trains' will be implemented by using Docker containers[12]. Docker containers have similar resource isolation and allocation benefits to virtual machines, creating temporary and secure sandboxes. However, Docker containers are more efficient and flexible as they virtualize the operation system rather than hardware.

The 'trains' sent to two PHT stations of data providers are designed for extracting and preparing data, pseudonymizing personal identifiers, encrypting the entire dataset, and sending it to the TSE station. One-way hashing function and random salts are shared by two data providers to make personal identifiers pseudonymized on both sites. Usually in cryptography, one-way hashing turns any format of data into a fixed-length "fingerprint" that cannot be reversed. Salt, as a random string, is appended to data before hashing, to eliminate the risk of malicious decryption. The pseudonymized personal identifiers are not related to a specific person, but shared hashing function and salts make it possible to link two datasets. The datasets including pseudonymized identifiers will be encrypted by public key to ensure safe transportation between data stations and the TSE station. When PHT data stations have sent their encrypted dataset to the TSE, data stations will produce a positive response (e.g., the message: 'OK') for indicating the 'train' has operated successfully.

After all encrypted data sets arrive at the TSE station, the researchers can trigger the application's execution including analytical algorithms and keys at the TSE. Within the application, there is one private key per PHT data station to decrypt the dataset, and one verification key to test the dataset integrity. In this case, the data provider station can only encrypt by using the public key, not decrypt, while only the TSE station maintains the private key to decrypt for this specific data provider. After the verification-decryption-verification process, data from both entities can be linked and merged by using Probabilistic Record Linkage [13, 14], i.e., records of CBS and the Maastricht Study of the same individual are merged. As the hashing function and salt, performed at the data station, are unknown to the TSE, it is not able to reverse or decrypt personal identifiers. In addition to use techniques of pseudonymization and encryption, a supplementary to preserve the privacy is that neither data providers have access to the TSE. After executing the analytical algorithms on the merged dataset, the TSE must validate the results which cannot reveal any identifiable information. Only the validated results which can contain multiple figures or tables can be sent back to researchers. Furthermore, all received and created datasets are deleted at the TSE.

## 3. PRELIMINARY RESULTS

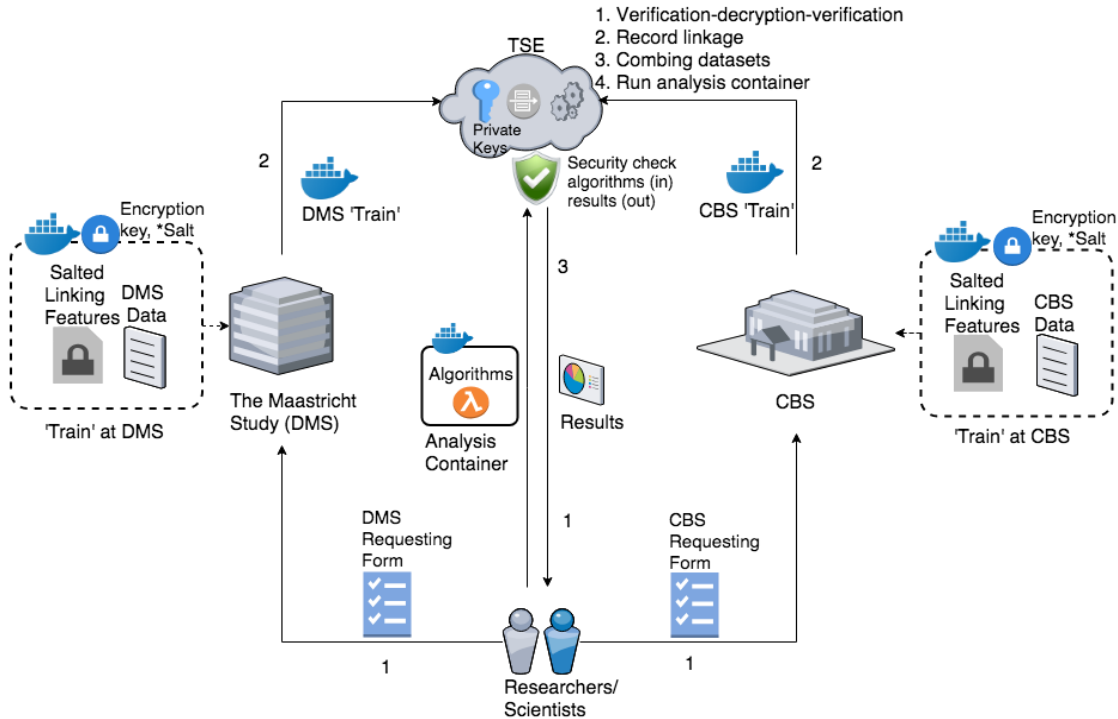

*Figure 1. Conceptual overview of the proposed infrastructure. The encryption and pseudonymization methods are implemented in the 'trains'. Every time data providers operate the 'train', different encryption keys will be generated at each site, while one salt needs to be created and agreed by both.*

In Figure 1, a conceptual overview of our proposed infrastructure is presented. Note that, part of this infrastructure has already been implemented and tested. First, an agreement was reached between two data providers on which features are available on both sites and suitable to link the datasets, while taking into account the legal constraints and the completeness and accuracy of the features. Four personal identifiers were selected to link two datasets: zip code, house number, gender, and date of birth. Second, we implemented the trains sent to data provider stations with Secure Hash Algorithm 2 (SHA-512) as the hashing function, random salt, and the public encryption key. The train for the TSE station was created with a simple question to plot two variables in one figure as the output, and private decryption key in place. Moreover, two data provider stations were built up with the one simulated personal identifier and one variable of the same size of instances. In the developing TSE station, verification, decryption by the private key, merging two datasets, execution of simple algorithms have been achieved. However, checking algorithmic processing credentials and validation of the final analysis results are still at the developing stage.

So far, this infrastructure has been tested successfully within one local machine in a simple simulated scenario where one data station holds unique ID with age, the other one holds the same unique ID with income. In the testing, after setting up all PHT stations, we only played the role of researchers who sent the data request and algorithm 'trains' to PHT stations. As long as the TSE station got the correct verification key, the result which was the figure with the relationship between age and income was sent back to researchers. Although the simulated scenario and data problem were very simple, it proves the feasibility of the whole infrastructure. At the current stage, to transform the infrastructure from simulated to a practical environment, the technical WP is installing PHT stations at CBS and UM sites respectively. The scientific WP is requesting data from two data sites 1) to answer the scientific questions, regarding the associations between lifestyle and health status, and 2) to evaluate the performance and security of our infrastructure.

## 4. CONCLUSION

This unique interdisciplinary project combines various scientific, technical, ethical, legal, and social challenges across different departments/organizations. At the management level, much effort is devoted to establishing an inspiring environment where three highly interlocked WPs can efficiently and effectively collaborate. Besides working within their own expertise, each WP is also responsible to maintain, update, and understandably explain their contributions to other WPs regularly.

Developing and installing a new infrastructure is not only a technical challenge, but it also involved exiting administrative standards, regulations, and policies from the receiving organizations. Although the infrastructure is tested using a simple simulated dataset, additional security enhancements and operational stability are required before the infrastructure is suited to deal with real (sensitive) data. Additionally, our test assumed that researchers are able to develop analysis algorithms without accessing the data directly. However, it is questionable whether this assumption holds in the context of real data. The infrastructure has been developed and will be improved to meet the requirements of security, accuracy, scalability, feasibility, sustainability, computation, labor, and time cost. Therefore, it is not possible to invent a perfect infrastructure which can absolutely cover all aspects, but we can find the best balance which is a perfect fit for current research requirements to use data.

## 5. REFERENCES

[1] Elshazly H., Azar A. T., El-korany A., Hassanien A. E. Hybrid system for lymphatic diseases diagnosis. Proceedings of the International Conference on Advances in Computing, Communications and Informatics (ICACCI '13); Mysore, India. IEEE; pp. 343–347. 2013.
[2] Dougherty G. Digital Image Processing for Medical Applications. Cambridge University Press; 2009.
[3] Tsymbal A., Meissner E., Kelm M., Kramer M. Towards cloud-based image-integrated similarity search in big data. Proceedings of the IEEE-EMBS International Conference on Biomedical and Health Informatics (BHI '14); June 2014; Valencia, Spain. IEEE; pp. 593–596.
[4] Chen W., Cockrell C., Ward K. R., Najarian K. Intracranial pressure level prediction in traumatic brain injury by extracting features from multiple sources and using machine learning methods. Proceedings of the IEEE International Conference on Bioinformatics and Biomedicine (BIBM '10); December 2010; IEEE; pp. 510–515.
[5]The determinants of health. (2018). Retrieved from http://www.who.int/hia/evidence/doh/en/
[6] S. Hardy et al., "Private federated learning on vertically partitioned data via entity resolution and additively homomorphic encryption," arXiv:1711.10677 [cs], Nov. 2017.
[7]M. D. Wilkinson *et al.*, "The FAIR Guiding Principles for scientific data management and stewardship," *Scientific Data*, vol. 3, p. 160018, Mar. 2016.
[8] M. Lawler et al., "A roadmap for restoring trust in Big Data," The Lancet Oncology, vol. 19, no. 8, pp. 1014–1015, Aug. 2018.
[9] Tal. Z. Zarsky, 'Incompatible: The GDPR in the Age of Big Data', 47 Seton Hall L. Rev. 995. 2017.
[10] Mireille Hildebrandt, Smart technologies and the end(s) of law: Novel entanglements of law and technology, 205. 2015.
[11]van S. Johan et al., "Using the Personal Health Train for Automated and Privacy-Preserving Analytics on Vertically Partitioned Data," Studies in Health Technology and Informatics, pp. 581–585, 2018.
[12]What is a Container - A standardized unit of software(2018). Retrieved from https://www.docker.com/resources/what-container
[13] I. P. Fellegi and A. B. Sunter, "A Theory for Record Linkage," p. 29.
[14] G. Kim and R. Chambers, "Regression Analysis under Probabilistic Multi-Linkage: *Regression analysis under probabilistic multi-linkage*," *Statistica Neerlandica*, vol. 66, no. 1, pp. 64–79, Feb. 2012.